\documentclass[conference]{IEEEtran}
\IEEEoverridecommandlockouts
\usepackage{cite}
\usepackage{amsmath,amssymb,amsfonts}
\usepackage{algorithmic}
\usepackage{graphicx}
\usepackage{textcomp}
\usepackage{xcolor}
\usepackage[table]{xcolor}
\def\BibTeX{{\rm B\kern-.05em{\sc i\kern-.025em b}\kern-.08em
    T\kern-.1667em\lower.7ex\hbox{E}\kern-.125emX}}

\usepackage{multirow}
\usepackage{multicol}
\usepackage{booktabs}
\usepackage{makecell}
\usepackage{adjustbox}
\usepackage{placeins}
\usepackage{threeparttable}
\usepackage{verbatim}
\usepackage{amsmath,amsfonts,bm}

\def\eqref#1{equation~\ref{#1}}
\def\1{\bm{1}}

\DeclareMathAlphabet{\mathsfit}{\encodingdefault}{\sfdefault}{m}{sl}
\SetMathAlphabet{\mathsfit}{bold}{\encodingdefault}{\sfdefault}{bx}{n}

\usepackage{amsthm}
\usepackage[ruled,linesnumbered]{algorithm2e}
\usepackage{tcolorbox}
\usepackage{listings}
\usepackage{makecell}

\usepackage{array}
\usepackage{amssymb}

\usepackage{caption}
\usepackage{array}
\newcolumntype{C}[1]{>{\centering\arraybackslash}m{#1}}

\renewcommand{\thesection}{\arabic{section}}

\begin{document}

\title{
Extreme-Scale Atomistic Simulation of Real-Temperature Magnetic Skyrmion Dynamics by Coupled Spin-Lattice Modeling
%Device-Scale Skyrmion Dynamics from First Principles with 100-billion-Atom Machine-Learning Spin–Lattice Simulation
}
%Device-Scale Skyrmion Dynamics from First Principles:\\Machine-Learning Spin--Lattice Simulation of 10 Trillion Atoms}
%on 12 Million ARM Cores}

\author{
\IEEEauthorblockN{
Pin Chen\textsuperscript{1},
Cheng-bing Chen\textsuperscript{2},
Hai Liu\textsuperscript{1},
Yuewen Huang\textsuperscript{1},
Kangyou Zhong\textsuperscript{1},
Hai-Jun Zhao\textsuperscript{3}, \\
Liu-Liu Han\textsuperscript{4,5},
Guixin Guo\textsuperscript{1},
Jiang Li\textsuperscript{1},
Dan Huang\textsuperscript{1},
Ben Xu\textsuperscript{2,*},
Yutong Lu\textsuperscript{1,6,*}
}
\IEEEauthorblockA{
\textsuperscript{1}Sun Yat-sen University, National Supercomputer Center in Guangzhou, Guangzhou, China \\
\textsuperscript{2}Graduate School of China Academy of Engineering Physics, Beijing 100089, China \\
\textsuperscript{3}Key Laboratory of Quantum Materials and Devices (MOE), \\School of Physics, Southeast University, Nanjing 211189, China \\
\textsuperscript{4}Suzhou Laboratory, Suzhou 215000, China \\
\textsuperscript{5}State Key Laboratory of Powder Metallurgy, Central South University, Changsha 410083, China \\
\textsuperscript{6}Sun Yat-sen University, National Supercomputer Center in Shenzhen, Shenzhen, China \\[4pt]
\textsuperscript{*}Corresponding authors: bxu@gscaep.ac.cn, luyutong@mail.sysu.edu.cn
}
}

\maketitle

% ====================================================================
% ABSTRACT — GB requirement: 150 word max
% ====================================================================
\begin{abstract}
Real-temperature topological magnetic dynamics in functional materials is governed by coupled lattice and spin evolution, yet remains inaccessible to predictive simulation at device-relevant scales. As a flagship example, thermally driven helix-to-skyrmion transformation in FeGe requires atomistic resolution, explicit lattice motion, and micrometer-scale domains to resolve device-scale topological texture formation. We combine a spin-constrained density-functional-theory-trained neuro-evolution potential with a structure-preserving spin--lattice integrator within one machine-learned framework. 
%Previous spin-aware atomistic simulations remain far smaller and rarely report whole-application deployability metrics. 
 Architecture--specific optimizations, kernel fusion, SVE2 vectorization, and NUMA-aware data layout deliver a seven orders-of-magnitude speedup over prior spin-aware methods.
 %deliver a $2.36\times$ speedup over baseline implementation.
Deployed on 
%a many-core ARM 
 LineShine exascale supercomputer, the full application scales to 12.45 million CPU cores with 89.7\% weak-scaling efficiency, enabling simulations of 1.34 trillion atoms and an equal number of spins while reaching 48.5 PFLOPS in double precision.
The simulations directly resolve real-temperature skyrmion nucleation and reorganization at previously inaccessible scales, establishing a new regime for predictive simulation of coupled spin--lattice topological magnetic dynamics.
\end{abstract}

%ARM SVE, Gordon Bell

% ====================================================================
% 1. JUSTIFICATION — GB requirement: 50 word max;
%    indicate what implementation or performance "high watermark" was
%    achieved (rather than the science that was enabled)
% ====================================================================
\section{Justification for the ACM Gordon Bell Prize}

We performed the largest atomistic simulation of a magnetic material to date: a 1.34-trillion spins and 1.34-trillion atoms coupled spin–lattice system executed on 12.45 million ARM CPU cores, achieving first-principles accuracy, reaching a sustained performance of 48.5 PFLOPS in double precision, and sustaining 89.7\% weak-scaling efficiency at full system scale. This unprecedented scale enables the first direct atomistic observation of skyrmion dynamics at device-relevant length scales.

% ====================================================================
% 2. PERFORMANCE ATTRIBUTES — GB requirement: table listing each
%    attribute title and value in a separate row
% ====================================================================
\section{Performance Attributes}

\begin{table}[!ht]
\centering
\label{tab:application_summary}
\rowcolors{2}{gray!15}{white}
\begin{tabular}{@{}ll@{}}
\toprule
\textbf{Performance attribute} & \textbf{This submission} \\
\midrule
Category of achievement & Peak performance, scalability \\
Type of method used & Explicit \\
Results reported on basis of & Whole application \\
Precision reported & Double precision \\
System scale & Full-scale machine (20{,}480 nodes) \\
Measurement mechanism & Timers and FLOP counting \\
\bottomrule
\end{tabular}
\end{table}

\FloatBarrier
% ====================================================================
% 3. OVERVIEW OF THE PROBLEM — GB requirement: 1 page max;
%    description of the problem and its importance, in terms
%    understandable to a non-specialist
% ====================================================================
\section{Overview of the Problem}
Magnetic skyrmions have emerged as one of the most compelling topological spin textures for future low-power information technologies. This is because they can combine nanoscale or mesoscale manipulability with nontrivial topology, therefore enabling rich dynamical functionality.  More broadly, Nobel Prize–recognized advances in spin-dependent transport and topological phases have established spin functionality and topology as two defining themes of modern condensed-matter physics~\cite{fert2008nobel,hasan2010colloquium}.

\begin{figure*}[htp]
    \centering
    \includegraphics[width=0.99\linewidth]{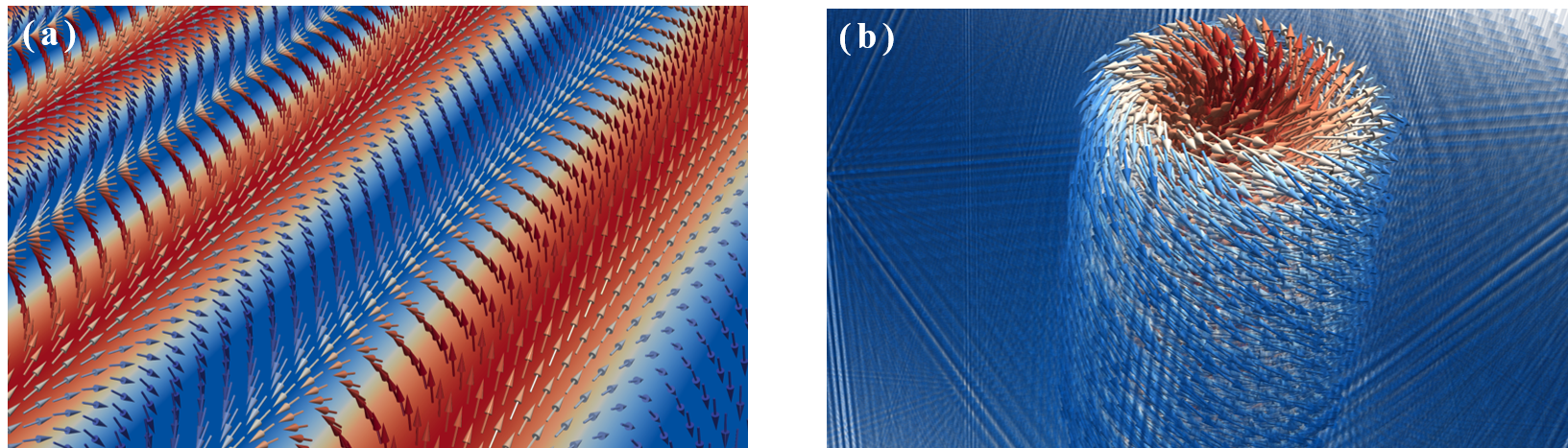}
    \caption{Schematic of a magnetic helix (a) and skyrmion (b) in a chiral magnet. Color coding represents the z component of magnetization. Arrows are plotted for every 200 atoms; the typical helix period and radius of skyrmion in FeGe is $\sim$70–80 nm~\cite{yu2011near}. 
    % \textcolor{red}{[Figures a and b need to be explained separately in the title]}
    }
    \label{fig:skyrmion_schematic}
\end{figure*}

Among the known skyrmion-hosting material classes, noncentrosymmetric chiral magnets provide one of the most important platforms because the competition between Heisenberg exchange(J) and the Dzyaloshinskii-Moriya interaction(D) naturally sets the characteristic magnetic length scale in the experimentally useful tens-of-nanometers range~\cite{bak1980helical,nagaosa2013topological}. Additionally, the three-dimensional nature of bulk DMI materials enables hosting of skyrmion strings and more complex topological particles like hopfions, offering richer physics and greater computational complexity than 2D systems. FeGe is a prototypical example: its characteristic helix/skyrmion size is about 70--80 nm~\cite{yu2011fege}, while its bulk critical helimagnetic ordering temperature lies close to room temperature, around 278 K, with a more precise single-crystal determination giving 279.1\,K~\cite{xu2017curie}. These two features make FeGe especially attractive for studying real-temperature skyrmion formation and dynamics.

However, these same features also make predictive simulation exceptionally demanding. First, a physically meaningful calculation must extend far beyond a single texture unit and instead contain many competing periods, boundaries, and fluctuation--mediated nucleation channels, which immediately push the system size toward the mesoscopic regime. Second, because the transformation occurs near experimentally relevant temperatures, temperature cannot be treated as a purely phenomenological parameter: thermal fluctuation, lattice motion, and spin--lattice energy exchange must be resolved explicitly. Third, the characteristic length scale is controlled by the delicate balance between exchange and DMI (J/D), so even modest errors in the effective J/D ratio can shift the helix pitch, skyrmion size, and phase competition. Finally, the relevant evolution unfolds on nanosecond time scales, which makes brute-force high-fidelity simulation prohibitively expensive unless both the potential evaluation and the coupled time integration can be executed at extreme throughput. Therefore, the evolution from a low-temperature helix-dominated state toward thermally stabilized skyrmionic textures in FeGe represents a central but still incompletely resolved problem~\cite{yu2010real,nagaosa2013topological}.

The central challenge is therefore not simply to simulate a larger or hotter magnetic system, but to resolve, \textbf{within one physically consistent framework, how real-temperature lattice motion and spin dynamics jointly drive the transformation from helical order to skyrmionic textures and their subsequent reorganization on experimentally relevant scales.}

%这个部分是否需要加一个skyrmions的示意图，可能计算机专家甚至部分做材料的专家都不了解这个？
%\subsection{Why skyrmions matter}

%\subsection{The scale challenge}

%The central difficulty of atomistic skyrmion simulation arises from the large separation of spatial scales involved in the underlying physics. Magnetic interactions such as exchange coupling, Dzyaloshinskii--Moriya interaction (DMI), and magnetocrystalline anisotropy originate from electronic structure and therefore must be resolved at atomic length scales on the order of $\sim$0.3\,nm~\cite{nagaosa2013topological}. In contrast, the nucleation and stability of skyrmions emerge from collective spin interactions spanning magnetic domains of tens to hundreds of nanometers, which correspond to experimentally relevant device dimensions~\cite{nagaosa2013topological}.

%Resolving these phenomena within an atomistic framework requires simulations containing tens to hundreds of billions of atoms, far beyond the scale accessible to existing atomistic magnetic simulations. Bridging this gap while explicitly resolving both spin and lattice degrees of freedom constitutes the primary computational challenge addressed in this work.

We address this challenge through three key components:
(1)~a magnetic machine-learning interatomic potential based on an extended spin-included neuro-evolution potential (NEP-SPIN), developed from the standard NEP \cite{fan2021neuroevolution}, which learns the spin-lattice coupled potential energy surface from constrained density functional theory data of magnetic excite configurations
%\cite{cai2022deltaspin,zheng2026integrating},
\cite{zheng2026integrating},
and replaces repeated electronic-structure evaluations with local atomic inference, reducing the computational cost from cubic-scaling first-principles calculations to overall linear scaling with system size.
%and replaces \textcolor{red}{need to check} the $O(N^3)$ electronic-structure calculation with an $O(N)$ local inference at each atom;
%(2)~The spin-lattice dynamics integrator designed for the magnetic machine learning library of interatomic potential(MMLIP) that propagates the coupled lattice and spin degrees of freedom with long-time numerical stability, including the longitudinal fluctuation of magnetic moment; 
(2) a symplectic spin-lattice dynamics integrator designed for magnetic machine learning library of interatomic potential (MMLIP) that propagates the coupled lattice and spin degrees of freedom with long-time numerical stability, and including the longitudinal fluctuation of magnetic moment
and (3)~an extreme-scale implementation of spin-lattice dynamics (termed magnetic molecular dynamics) within LAMMPS optimized for a many-core ARM supercomputer, enabling simulations on 20{,}480 nodes (12.45 million CPU cores). Our method enables atomistic simulations of up to $1.34\times10^{12}$ magnetic atoms, corresponding to a physical domain of $3.02\,\mu\mathrm{m} \times 2.41\,\mu\mathrm{m} \times 2.41\,\mu\mathrm{m}$ and reaching experimentally relevant device dimensions. In the present model, each atom carries one local magnetic moment, so the number of spins is equal to the number of atoms. Throughout the paper, system size is reported by atom count unless otherwise noted.

% ====================================================================
% 4. CURRENT STATE OF THE ART — GB requirement: 1 page max;
%    quantitative discussion of current SoA, with emphasis on
%    performance-related aspects
% ====================================================================
\section{Current State of the Art}

\begin{table*}[t]
\centering
\caption{Performance comparison of representative atomistic and spin--lattice simulation frameworks at scale.
$\uparrow$\,= higher is better; $\downarrow$\,= lower is better.
Abbreviations: D.O.F.\ = degrees of freedom; TtS = time-to-solution; A = atom-only; S{+}A = coupled spin--lattice;
Norm.\ TtS = TtS normalized by model parameter count, i.e.\ s/(atom$\cdot$parameter$\cdot$step).
$^{*}$Atom-only method without magnetic degrees of freedom; included as Gordon Bell performance baselines.
$^\dagger$Projected from 1/16 machine (9{,}936 nodes) to full machine; see~\cite{guo2022extending} for details.
$^\ddagger$Mixed precision (FP64/FP32/BF16); not a pure FP64 result. All other sustained FLOPS entries are double precision.}
\label{tab:sota_spin_lattice}
\scriptsize
\setlength{\tabcolsep}{4pt}
\renewcommand{\arraystretch}{1.20}
\begin{adjustbox}{max width=\textwidth}
\begin{tabular}{
l
c c c c
c c c
c c c c
}

\toprule
\textbf{Work} &
\textbf{Year} &
\textbf{Method} &
\textbf{System} &
\textbf{D.O.F.} &
\makecell[c]{\textbf{\# atoms/}\\\textbf{spins}} &
\makecell[c]{\textbf{\# CPU}\\\textbf{cores}} &
\makecell[c]{\textbf{\#}\\\textbf{GPUs}} &
\textbf{Machine} &
\makecell[c]{\textbf{Sustained}\\\textbf{[FLOPS]\,$\uparrow$}} &
\makecell[c]{\textbf{TtS\,$\downarrow$}\\\textbf{[s/step/atom]}} &
\makecell[c]{\textbf{Norm.\ TtS\,$\downarrow$}\\\textbf{[s/(atom$\cdot$param}\\\textbf{$\cdot$step)]}} \\
\midrule

Baseline~\cite{jia2020pushing}$^{*}$(double)
  & 2020 & DeepMD & Cu & A
  & 127\,M & 27.3\,K & 27.3\,K & Summit & 91\,P & $8.1\times10^{-10}$ & --- \\

Baseline~\cite{jia2020pushing}$^{*}$(double)
  & 2020 & DeepMD & H$_2$O & A
  & 679\,M & 27.3\,K & 27.3\,K & Summit & 80\,P & $8.1\times10^{-10}$ & --- \\

Guo et al.~\cite{guo2022extending}$^{*}$(double)
  & 2022 & DeepMD & Cu & A
  & 3.4\,B & 27.3\,K & 27.3\,K & Summit & 43.7\,P & $1.1\times10^{-10}$ & --- \\

Guo et al.~\cite{guo2022extending}$^{*}$(double)
  & 2022 & DeepMD & H$_2$O & A
  & 3.9\,B & 27.3\,K & 27.3\,K & Summit & 46.3\,P & --- & --- \\

Guo et al.~\cite{guo2022extending}$^{*\dagger}$(double)
  & 2022 & DeepMD & Cu & A
  & 17.3\,B & 7.63\,M & --- & Fugaku & 119\,P & $4.1\times10^{-11}$ & --- \\

Guo et al.~\cite{guo2022extending}$^{*\dagger}$(double)
  & 2022 & DeepMD & H$_2$O & A
  & 24.9\,B & 7.63\,M & --- & Fugaku & 124.8\,P & --- & --- \\

Razakh et al.~\cite{xsnnqmd25}$^{*}$ (mixed)
  & 2025 & XS-NNQMD & PbTiO$_3$ & A
  & 1.23\,T & 1.04\,M & 60\,K & Aurora & 1.87\,E$^\ddagger$ & --- & $1.876\times10^{-15}$ \\

\midrule

Dudarev et al.~\cite{spilady08}
  & 2008 & Spilady & Fe & S{+}A
  & 128\,K & --- & --- & --- & --- & --- & --- \\

Yu et al.~\cite{spingnn+24}
  & 2024 & spinGNN++ & --- & S{+}A
  & 4{,}608 & --- & 16 & --- & --- & --- & --- \\

Yang et al.~\cite{deepspin24}
  & 2024 & DeepSPIN & --- & S{+}A
  & 11{,}328 & 5.6k & --- & --- & --- & $4.52\times10^{-3}$ & --- \\

\midrule
\textbf{This work}
  & 2026 & NEP + SPIN & FeGe & S{+}A
  & \makecell{168\,B \\ \textbf{1.34\,T}}
  & 12.45\,M & --- & LineShine
  & \makecell{48.5\,P \\ 43.3\,P}
  & \makecell{$\mathbf{1.79\times10^{-11}}$ \\ $1.82\times10^{-11}$}
  & \makecell{$3.97\times10^{-15}$ \\ $4.05\times10^{-15}$} \\

\bottomrule
\end{tabular}
\end{adjustbox}
\end{table*}

State-of-the-art simulation of thermally driven topological magnetic phenomena requires a unified spin–lattice framework that treats atomic positions and local magnetic moments on an equal footing, enabling consistent computation of forces and magnetic torques for time integration. This requirement is particularly critical for modeling skyrmion formation, where the underlying physics spans multiple scales: from atom-resolved lattice distortions and texture reorganization at 10–100 nm, to collective dynamics across mesoscopic or even micrometer-sized domains over extended timescales. In the context of Gordon Bell-class computing, these scientific demands translate into stringent system requirements: the online evaluation of spin--lattice dynamics must be local, massively parallel, communication-efficient, and accelerator-friendly, or production-scale simulation of real-temperature topological nucleation remains out of reach.
Table~\ref{tab:sota_spin_lattice} summarizes representative approaches across this landscape, including atom-only ML-MD baselines from the 2020 Gordon Bell Prize (DeepMD~\cite{jia2020pushing}) and the 2025 Gordon Bell finalist (XS-NNQMD~\cite{xsnnqmd25}), existing spin--lattice methods, and the present work.

In this perspective, existing methods still trade physical fidelity with computational scale. Continuum micromagnetics can reach large domains, but temperature usually enters only through effective noise or fitted parameters, so lattice heating and lattice-to-spin energy transfer are not treated explicitly\cite{chubykalo2006dynamic}. Atomistic spin dynamics resolves individual moments, yet the lattice is often frozen or replaced by a thermal bath\cite{evans2014atomistic}. A more complete route is on-the-fly first-principles spin--lattice dynamics. In practice, however, its cost is set not only by electronic-structure evaluations, but also by the need to follow nonadiabatic transitions among multiple electronic or spin states, often in a surface-hopping-type framework\cite{tully1990molecular,crespo2018nonadiabatic}. More scalable alternatives include DFT-parameterized spin Hamiltonians, spin- or magnetic-cluster expansions, and classical spin--lattice dynamics\cite{he2021tb2j,mace24,lavrentiev2010magnetic,spilady08,tranchida18}. These methods are efficient and interpretable, but their magnetic couplings are usually fixed in advance rather than updated self-consistently with temperature-driven atomic motion. This limits transferability when the structure changes strongly, the pathway is complex, or real-temperature evolution is essential.

Recent spin-aware machine-learned models close this gap by describing energy, force, and magnetic torque within a unified representation, through explicit incorporation of magnetic environments into the descriptor or neural architecture. This category includes spin--lattice hybrid ML potentials\cite{mslp22}, magnetic atomic cluster expansion (ACE) variants\cite{mace24}, graph-based magnetic networks\cite{spingnn24,spingnn+24}, and other torque-aware models\cite{deepspin24,mMTP22,mMTP23} developed for non-collinear and finite-temperature settings. Their promise is substantial: they offer a route toward combining anharmonic lattice dynamics, complex magnetic interactions, and local spin--lattice feedback within a single differentiable surrogate.

The remaining difficulty, however, is now less conceptual than computational. At inference of spin and lattice configuration and energy, the practical cost depends strongly on descriptor locality, kernel fusion, memory-access efficiency, and hardware suitability.
%CPU/GPU suitability. 
In this regard, models such as DeepSPIN\cite{deepspin24} represent an important step toward explicit spin-aware learning, but implementations based on general-purpose deep-learning frameworks may introduce substantial inference overhead in long-time, large-system simulations. By contrast, NEP-style local descriptors\cite{NEP21} are much closer to the throughput and scalability requirements of production HPC workflows, but in their current form they do not explicitly include spin degrees of freedom or generate magnetic torques.

%这句话看起来很重要，可以加粗。 
\textbf{The central state-of-the-art gap is therefore not simply one of accuracy or scale in isolation, but of closing physical completeness and exascale deployability within the same online potential-evaluation pipeline}. Among the currently available spin-aware ML approaches, SpinGNN\cite{spingnn24} explicitly reports million-atom GPU-parallel spin--lattice simulations, whereas DeepSPIN and SpinGNN++\cite{deepspin24,spingnn+24} place stronger emphasis on physical fidelity and model expressiveness. Publicly available papers still rarely report Gordon-Bell-style metrics such as peak FLOPS or seconds per step per atom, which makes their exascale deployability difficult to assess directly; as shown in Table~\ref{tab:sota_spin_lattice}, most spin--lattice methods lack reported performance metrics entirely.

This work targets precisely this gap.
By extending the NEP descriptor to incorporate non-collinear spin degrees of freedom, coupling it with a spin–lattice integrator designed to preserve the key geometric structure of the coupled dynamics and maintain long-time numerical stability, and co-designing the entire pipeline for a many-core ARM supercomputer, we achieve coupled spin--lattice simulations of up to 1.34\,T atoms at a time-to-solution of $1.79\times10^{-11}$\,s/step/atom in double precision.
%(``This work'' row in Table~\ref{tab:sota_spin_lattice}).
The following sections describe the algorithmic and implementation innovations that make this possible.

\section{Innovations Realized}
\label{sec:innovations}

\begin{figure*}
    \centering
\includegraphics[width=0.99\linewidth]{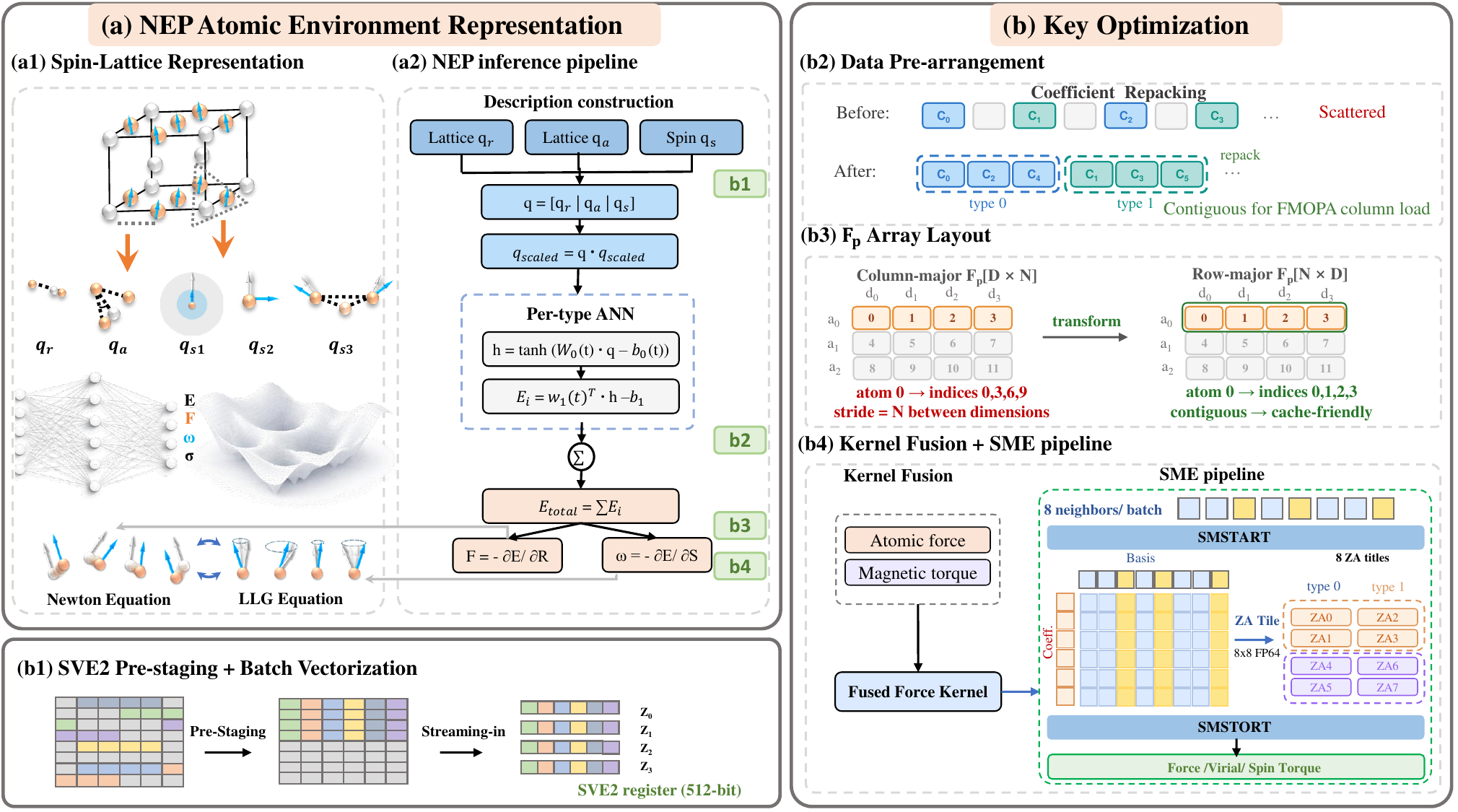}
    \caption{
    Fine-grained overview of the DFT-trained NEPSPIN framework for spin-lattice molecular dynamics. (a) Atomic-environment representation and NEP inference workflow for spin-lattice interactions. The workflow in panel (a2) is partitioned into four highlighted stages, denoted b1--b4. 
    (b) Performance-oriented implementation optimizations corresponding one-to-one to these four stages in (a2), indicating precisely where data pre-arrangement, array-layout transformation, SVE2 pre-staging and batch vectorization, and kernel fusion with SME execution are applied along the pipeline.    }
    \label{fig:framework}
\end{figure*}

Porting NEPSPIN-based spin--lattice dynamics to the target many-core ARM platform exposes several challenges rooted in the mismatch between the model's computational patterns and the hardware micro-architecture: the deep 16-domain NUMA hierarchy demands careful affinity management; the absence of a shared L3 cache penalizes data reuse across memory-bound descriptor kernels; and the dominance of GEMV-like inference operations underutilizes the SME matrix units designed for outer-product GEMM workloads.
% Addressing these challenges motivates the algorithmic and implementation innovations described below.
Addressing these challenges motivates the algorithmic and implementation innovations described in the following and illustrated in Fig.~\ref{fig:framework}.

\subsection{Algorithmic Innovations}
\subsubsection{A spin-aware local descriptor pipeline}

We extended the existing NEP inference path~\cite{fan2021neuroevolution} to incorporate magnetic information while preserving the same locality, cutoff-based neighbor traversal, and regular evaluation pattern as in the structural descriptor pipeline. In practice, the local feature vector augments the original structural channels with three groups of magnetic channels that correspond to onsite, pairwise, and angular spin-lattice information. This organization allows the magnetic extension to reuse the same radial-basis and angular-accumulation infrastructure as the structural part, keeping the descriptor evaluation local and implementation-friendly.

\subsubsection{Implementation-oriented descriptor organization}

The magnetic channels are organized in three groups. The first group collects onsite information from the local spin state. The second group evaluates pairwise spin-bond couplings over the neighbor list using the same radial carrier as the structural interaction channels. The third group evaluates spin-weighted angular information by accumulating directional channels over neighbors and contracting them into rotationally invariant quantities. Additional mixed channels are formed through local contractions of these accumulated quantities. From an implementation perspective, all magnetic channels follow the same basic computational pattern: local neighbor traversal, channel-wise accumulation, and small dense contractions. As a result, the magnetic extension increases arithmetic work without introducing new global data dependencies or irregular communication paths.

\subsubsection{Self-consistent midpoint spin update}

For cases with strong feedback between the spin state and the effective field, we augment the explicit predictor-corrector update with an optional self-consistent midpoint iteration. Starting from the beginning-of-step spin state, the algorithm repeatedly forms a midpoint configuration, reevaluates the force and effective field at that midpoint, and reapplies the existing one-step update until either a convergence criterion or an iteration cap is reached. An accelerated fixed-point variant with regularization can be enabled for strongly nonlinear cases. This preserves the original operator structure while improving robustness under strong state dependence. Because the procedure may trigger multiple force/field reevaluations within a single time step, the spin update must be scheduled last among time-integration operations.

% --- 5.2 Implementation innovations in extreme-scale LAMMPS deployment
\subsection{Implementation Innovations}
% \textcolor{red}{Here's a mark: too many details at the code level need to be masked. ..TBD..}

\subsubsection{Unified spin--lattice execution in LAMMPS}

The proposed framework is realized in LAMMPS by coupling the NEPSPIN-based force-field evaluation with the spin--lattice time integrator equipping the above self-consistent updating scheme. 
At each MD step, the NEPSPIN module performs descriptor construction and inference on the unified energy surface $E(\mathbf{R},\mathbf{S})$, where $\mathbf{R}$ and $\mathbf{S}$ denote the atomic positions and spin configurations respectively, producing both lattice and spin driving terms for the subsequent coupled update. 
The spin--lattice integrator then advances the state $(\mathbf{R},\mathbf{S})$ with atomic forces and magnetic torques in a single update procedure, followed by halo exchange for the next step. 
This implementation eliminates the need for separate lattice and magnetic solvers, and preserves consistency between force evaluation, spin evolution, and parallel data movement throughout the simulation.

% The proposed framework is realized in LAMMPS by coupling the NEPSPIN-based force-field evaluation with the spin--lattice time integrator equipping the above self-consistent updating scheme. 
% At each MD step, the NEPSPIN module performs descriptor construction and inference on the unified energy surface $E(\mathbf{R},\mathbf{S})$, producing both lattice and spin driving terms for the subsequent coupled update. 
% The spin--lattice integrator then advances the state $(\mathbf{R},\mathbf{S})$ with atomic force $\mathbf{F}$ and magnetic torque $\mathbf{\omega}$ in a single update procedure, followed by halo exchange for the next step. 
% This implementation eliminates the need for separate lattice and magnetic solvers, and preserves consistency between force evaluation, spin evolution, and parallel data movement throughout the simulation.

\subsubsection{Fused multi-physics force kernel}

In the original NEPSPIN implementation, forces and magnetic torques are evaluated by three independent functions, each performing a full neighbor-list traversal with its own distance computation, cutoff test, and Chebyshev basis recurrence.
Because all three share the same radial cutoff, we fuse them into a single kernel that walks the neighbor list once and evaluates all force and torque contributions per pair in a single pass.
This eliminates two redundant neighbor traversals (${\sim}300$ instructions per neighbor $\times$ ${\sim}110$ neighbors per atom) and two redundant Chebyshev recurrences (244 instructions per neighbor each), reducing the fused-kernel runtime by 23.9\%.

\subsubsection{SVE2 predicated vectorization with pre-staging}

The irregular neighbor list---where each atom has a different number of valid neighbors after cutoff filtering---prevents direct vectorization of the inner loop.
We address this with a two-phase \emph{pre-staging} strategy.
In Phase~A, a scalar pass filters valid neighbors through cutoff tests and packs their coordinates, distances, spins, and element types into a contiguous, \texttt{alignas(64)} SoA buffer.
In Phase~B, the packed buffer is processed in 8-wide SVE2 batches (matching the 512-bit SVL for FP64).
Three techniques are critical within the SVE2 kernel:
(i)~\emph{online Chebyshev recurrence}---basis functions $T_{k+1}=2xT_k-T_{k-1}$ are evaluated inside the vector register file via a running pair of SVE vectors, avoiding the sizeless-type array limitation of ARM SVE;
(ii)~\emph{predicated multi-type dispatch}---\texttt{svsel\_f64} selects per-lane coefficients for different element types (e.g.\ Fe vs.\ Ge) in a single instruction, eliminating branch divergence and type-sorted gather/scatter;
(iii)~\emph{gather-load coefficient inner products}---strided ANN coefficients are loaded via \texttt{svld1\_gather} and contracted with contiguous basis vectors in one fused multiply--horizontal-reduce sequence.
Together with a column-major-to-row-major transposition of the descriptor-derivative array \texttt{g\_Fp} (placing each atom's 66 Fp parameters in ${\sim}8$ contiguous cache lines), these optimizations deliver a combined 29.6\% reduction in loop time over the fused kernel alone (21.74\,s$\to$15.30\,s).

\subsubsection{SME three-stage pipeline with predicate-driven type disambiguation}

The most architecture-specific innovation is the reformulation of the dominant coefficient inner products as outer-product GEMM operations executed on the ARM SME matrix engine.
The NEPSPIN force kernel is restructured into a three-phase pipeline operating on batches of 8~neighbors.

In the \emph{preparation} phase, scalar distance computation, cutoff filtering, and Chebyshev basis recurrence produce per-neighbor basis vectors $\mathrm{fn}_{12}[k]$ and $\mathrm{dfn}_{12}[k]$, written into an SoA buffer with a $[\text{basis}][\text{batch}]$ layout so that the batch dimension is contiguous for FMOPA row operands.
Shared Chebyshev polynomials $T_k$, $U_k$ are computed once per neighbor pair and reused across all four magnetic sub-terms (Exchange, DMI, ANI, SIA), eliminating three redundant recurrences (${\sim}120$ FLOP per pair).
Precomputed products $\mathrm{fp}{\cdot}\mathrm{dC}$ and $\mathrm{fp}{\cdot}\mathrm{Cv}$ are factored out of the sub-term loops, reducing per-$k$ cost from 6 to 2~FLOP.

The \emph{SME GEMM} phase performs the coefficient--basis inner products that dominate the per-neighbor cost.
A mixed-type neighbor batch poses a data-layout challenge: in a na\"ive implementation, Fe and Ge neighbors must be separated by gather/scatter operations before their respective coefficient vectors can be contracted, adding 35\% overhead.
We eliminate this overhead through \emph{predicate-driven type disambiguation}.
An \texttt{\_\_arm\_locally\_streaming} function maps the eight available ZA Tiles ($8\times8$ FP64 each) into four logical groups: ZA0--ZA1 and ZA2--ZA3 accumulate radial coefficient--basis products ($\mathrm{C}_{\mathrm{rad}} \times \mathrm{dfn}/\mathrm{fn}$) for Fe and Ge respectively, while ZA4--ZA5 and ZA6--ZA7 accumulate the corresponding spin products ($\mathrm{C}_{\mathrm{spin}} \times \mathrm{dfn}/\mathrm{fn}$).
Within each FMOPA instruction, the column predicate is set according to the element type of each lane, masking inactive lanes to zero so that Fe and Ge contributions accumulate into their designated tiles without data reshuffling.
After the basis loop completes, the two per-type tiles in each group are reduced by element-wise addition.
For a basis size of 8 (9 recurrence iterations), the entire GEMM phase issues 126~instructions per 8-neighbor batch, equivalent to 1{,}008~FMA operations.

In the \emph{post-processing} phase, each neighbor's force and torque contributions are assembled from the GEMM results using the precomputed tables $\mathrm{fp}{\cdot}\mathrm{dC}$/$\mathrm{fp}{\cdot}\mathrm{Cv}$ and the cached geometric/spin quantities, with results accumulated into thread-private force buffers followed by a parallel reduction.

The three-stage pipeline delivers an additional 11.2\% loop-time reduction (13.63\,s$\to$12.11\,s) on top of the SVE2 optimizations.
Across all five architecture-specific optimizations, the cumulative single-node speedup is $2.36\times$ over the OpenMP baseline and $70.9\times$ over the serial code (Fig.~\ref{fig:ablation}).
Combined with MPI spatial decomposition across nodes and OpenMP thread parallelism within each rank, this three-level execution model enables the coupled spin--lattice application to scale to 20{,}480 nodes and 12.45 million CPU cores.

% ====================================================================
% 6. HOW PERFORMANCE WAS MEASURED — GB requirement:
%    (a) what application was used to measure performance (1 page max)
%    (b) system and environment where performance was measured (1 page max)
%    Preference given to performance actually measured (not projected),
%    based on entire application (including I/O), uniform precision.
% ====================================================================
\section{How Performance Was Measured}

% --- 6.1 System and environment where performance was measured
\subsection{System architecture and hardware platform}

\begin{figure}
    \centering
\includegraphics[width=0.9\linewidth]{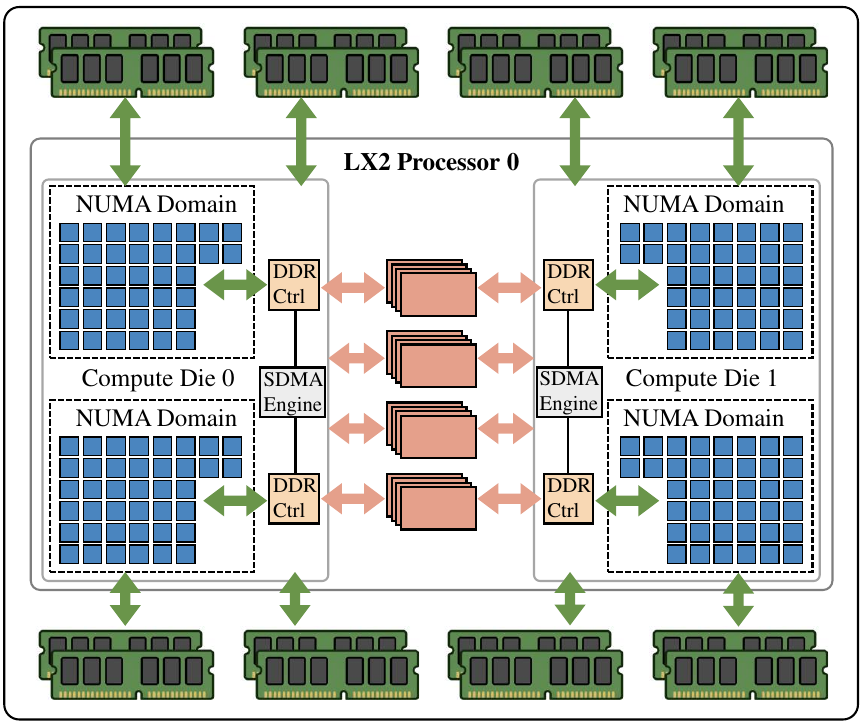}
    \caption{Internal architecture of an LX2 processor, comprising two compute dies, on-package HBM stacks, SDMA engines, and four DDR-attached NUMA domains.}
    \label{fig:cmputing_node}
\end{figure}

All measurements were performed on the LineShine supercomputer, a newly deployed exascale system unveiled in April 2026 by the National Supercomputing Center in Shenzhen (NSCC-SZ), China.
Each node contains two Armv9-based LX2 processors; the internal architecture of a single processor is illustrated in Fig.~\ref{fig:cmputing_node}.
The LX2 integrates two compute dies (304 cores total) and eight on-package HBM stacks (32\,GB, 4\,TB/s aggregate bandwidth) in a single package.
Each compute die contains 152 cores and is paired with 128\,GB of off-package DDR memory organized into four NUMA domains, for a total of 256\,GB DDR per processor.
Within each NUMA domain, cores share HBM; each compute die includes a dedicated SDMA engine for data movement between DDR and HBM.
At the node level, the two LX2 processors provide 608 physical cores, 64\,GB HBM, and 512\,GB DDR, arranged in 16 NUMA domains.
The memory subsystem has no shared last-level cache, making performance highly sensitive to data locality---a constraint addressed by the layout and pre-staging optimizations described in Section~5.
The complete per-node hardware specification is summarized in Table~\ref{tab:node_spec}.

The LX2 supports FP64, FP32, FP16, and INT8 through SME and SVE units, delivering up to 60.3\,TFLOP/s FP64 per processor (120.6\,TFLOP/s per node).
Large-scale compute nodes are interconnected via the LingQi high-speed network with a dual-plane, multi-rail fat-tree topology, providing 1.6\,Tb/s bandwidth per node.

\begin{table}[htp]
\centering
\caption{Per-node hardware specification.}
\label{tab:node_spec}
\begin{tabular}{ll}
\toprule
Attribute & Value \\
\midrule
Processor & Armv9-based LX2 ($\times$2 per node) \\
ISA & Armv9-A \\
Extensions & SVE2, SME \\
Cores per processor & 304 (2 dies $\times$ 152) \\
Physical cores per node & 608 \\
HBM per processor & 32\,GB (8 stacks, 4\,TB/s) \\
DDR per processor & 256\,GB (4 NUMA domains per die) \\
NUMA domains per node & 16 \\
DP peak per processor & 60.3\,TFLOP/s (SME) \\
Network bandwidth per node & 1.6\,Tb/s (LingQi fat-tree) \\
\bottomrule
\end{tabular}
\end{table}

The entire system delivers over 2.5 \,EFLOP/s peak performance in FP64.
Our experiments use up to 20{,}480 nodes (12.45 million cores), corresponding to full-scale machine allocation.

% --- 6.2 What application was used to measure performance
\subsection{Benchmark application and setup}
To measure performance, we used the real-temperature helix-to-skyrmion transformation in FeGe as a whole-application benchmark for coupled spin--lattice dynamics. This benchmark is particularly demanding because every stage of the simulation pipeline remains active throughout the transition, including communication, descriptor evaluation, force-and-torque inference, and coupled time integration. Our performance-optimized spin--lattice dynamics evolves coupled atomic and spin degrees of freedom for up to $1.34\times10^{12}$ atoms and an equal number of spins in a FeGe system of size $3018\times2414\times2414$\,\AA{} ($3.02\times2.41\times2.41$\,\textmu m). We first prepare a low-temperature helical state under experimentally relevant geometry and boundary conditions, and then drive the system through the temperature regime in which skyrmion seeds emerge, proliferate, and reorganize into extended topological textures. The measured workload therefore includes neighbor construction, halo exchange, spin-aware descriptor evaluation, unified force-and-torque inference, and long-time coupled time integration.

\begin{figure}[thp]
    \centering
    \includegraphics[width=0.99\linewidth]{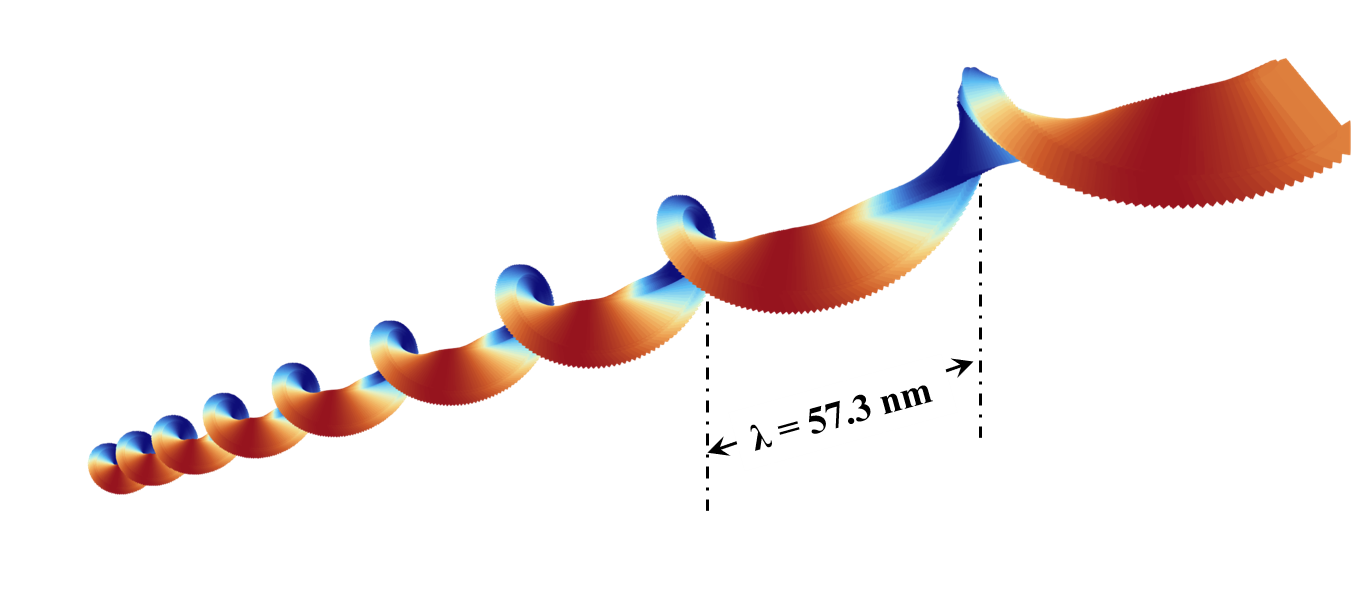}
    \caption{Schematic illustration of the helical magnetic structure in FeGe.
Arrows represent local spin orientations with a helical period of
$\lambda = 57.3$~nm. Colors indicate the $x$-component of the spin
$S_x$, ranging from blue ($S_x = -1$) to red ($S_x = +1$).}
    \label{fig:helix}
\end{figure}
Capturing this process requires substantially more than identifying equilibrium phases: the helix pitch must first be reproduced quantitatively, because it sets the characteristic magnetic length scale through the competition among exchange, Dzyaloshinskii–Moriya interaction, anisotropy, and Zeeman energy. As shown in Fig. ~\ref{fig:helix}, our model reproduces the zero-field helix pitch of FeGe over the temperature range of interest. 

Performance is characterized by five complementary metrics, summarized in Table~\ref{tab:metrics}.
All metrics are derived from a common timing primitive, the MD \texttt{Loop time}, defined as the wall-clock time spent in the time-stepping run loop, excluding one-time setup costs.
The application-level floating-point throughput (FLOPS) is estimated from the dominant NEPSPIN workload in each MD step, including radial and angular pair interactions, ANN evaluation, spin-dependent terms, and optional short-range interactions; the modeled FLOPs per step are summed across all MPI ranks and divided by the \texttt{Loop time}.
Different subsections of Section~\ref{sec:results} select the metric most suited to each analysis: loop time for the ablation study, atom-step/s for cross-framework comparison, and parallel efficiency, speedup, and sustained FLOPS for scalability characterization.

\begin{table}[htp]
\centering
\caption{Performance metrics used in this work.  All are derived from the
         MD \texttt{Loop time}; $N_{\mathrm{atom}}$ denotes the total atom
         count, $N_{\mathrm{step}}$ the number of time steps, and subscripts
         $1$ and $N$ refer to single-node and $N$-node measurements.}
\label{tab:metrics}
\begin{tabular}{ll}
\toprule
Metric & Definition \\
\midrule
Loop time (s)
  & wall-clock time of run loop \\
atom-step/s/node
  & $N_{\mathrm{atom}} \times N_{\mathrm{step}} / \text{Loop time}$ / $N_{\mathrm{node}}$ \\
Parallel eff.\ $\eta$
  & $T_{1}/T_{N}$ (weak) or $S/S_{\mathrm{ideal}}$ (strong) \\
Speedup
  & $T_{\mathrm{base}}/T_{N}$ \\
Sustained FLOPS
  & modeled FLOPs per step $\times\, N_{\mathrm{step}} / \text{Loop time}$ \\
\bottomrule
\end{tabular}
\end{table}

% ====================================================================
% 7. PERFORMANCE RESULTS — GB requirement: 2 pages max;
%    include scalability (weak and strong), time to solution,
%    efficiency (of bottleneck resources), and peak performance
% ====================================================================
\section{Performance Results}
\label{sec:results}

% --- 7.1 Single-node performance: ablation study + DeePMD comparison
\subsection{Single-node performance}

Before presenting multi-node scaling results, we establish single-node efficiency through two complementary analyses: an ablation study validating each architecture-specific optimization, and a direct comparison with DeePMD~\cite{jia2020pushing} on identical hardware.
DeePMD is chosen as the reference because it is the ML interatomic-potential framework featured in the 2020 Gordon Bell Prize and remains the most widely adopted baseline for large-scale ML-MD performance evaluation.
We compare the two frameworks in both prediction accuracy and single-node computational throughput to demonstrate that NEPSPIN achieves competitive physical fidelity while delivering substantially higher efficiency on the target ARM architecture.

\subsubsection{Optimization ablation study}
\label{sec:ablation}

\begin{figure}[htp]
    \centering
    \includegraphics[width=0.95\linewidth]{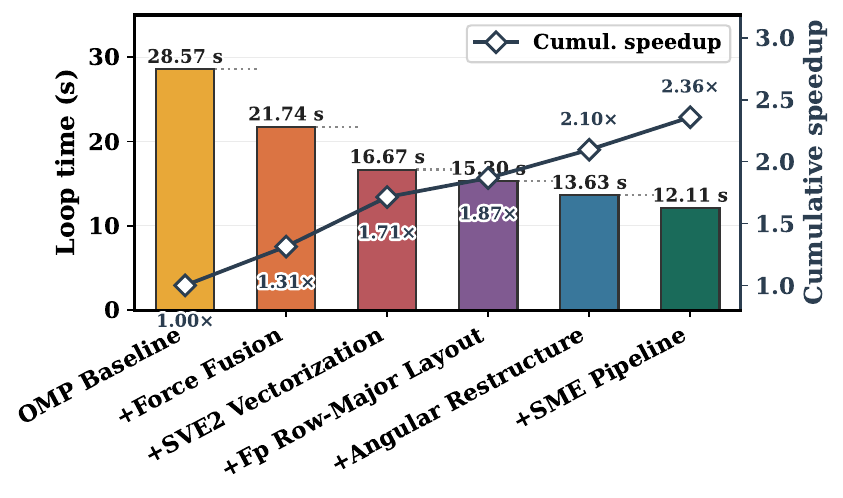}
    \caption{Ablation study of single-node optimizations.
    Each bar shows the loop time after cumulatively applying the listed optimization.
    Annotations indicate the incremental reduction from the previous step.
    Gray bar: OMP thread-parallel baseline; blue bars: architecture-specific optimizations.}
    \label{fig:ablation}
\end{figure}

Starting from a serial baseline of 858.04 s per loop (single-threaded NEPSPIN force evaluation at one node), OpenMP thread parallelism across 608~cores reduces the loop time to 28.57 s (30.0$\times$).
Then, five architecture-specific optimizations are applied cumulatively. Fig.~\ref{fig:ablation} reports the measured loop time after each step.

The first two optimizations yield the largest individual gains.
\emph{Spin-radial force fusion} (Step~1) merges three independent force kernels --radial, spin-lattice, and magnetic torque---into a single neighbor traversal, eliminating two redundant neighbor walks and two redundant Chebyshev basis evaluations per atom; this reduces loop time by 23.9\% (28.57 s$\to$21.74 s).
\emph{SVE2 vectorization} (Step~2) restructures the fused kernel into a batch processing pipeline that collects valid neighbors in contiguous SoA buffers and processes them in 8-wide SVE2 batches with predicated type-aware coefficient selection, delivering an additional 23.3\% reduction (21.74~s$\to$16.67~s).
The remaining three optimizations target data layout and instruction-level efficiency:
\emph{Fp row-major layout} (Step~3) transposes the descriptor-derivative array so that each atom's 66 Fp parameters occupy contiguous cache lines, improving L1 hit rate ($-$8.2\%);
\emph{angular descriptor restructure} (Step~4) hoists shared Chebyshev polynomials and precomputes fp$\cdot$dC / fp $\cdot$ Cv products across Exchange/DMI/ANI/SIA sub-terms, cutting redundant FLOPs ($-$11.0\%);
The \emph{SME three-stage pipeline} (Step~5) reformulates the coefficient inner products as outer-product GEMM operations using all eight ZA Tiles with predicate-driven type disambiguation, achieving a further 11.2\% reduction.
Cumulatively, the five architecture-specific optimizations achieve a $2.36\times$ speedup over the OMP baseline (28.57\,s$\to$12.11\,s), and the total speedup from serial is $70.9\times$.

\subsubsection{Comparison with DeePMD}
\label{sec:deepmd}

To contextualize NEPSPIN, we compare it against DeePMD~\cite{jia2020pushing}, the ML-potential framework featured in a prior Gordon Bell Prize, in both accuracy and computational throughput.

%\paragraph{Accuracy.}
%\noindent\textbf{
\textit{Accuracy:}
%}
Table~\ref{tab:accuracy_comparison} compares the prediction errors of NEPSPIN and DeePMD on the same FeGe spin--lattice validation set derived from spin-constrained DFT calculations.
NEPSPIN achieves an energy RMSE of 1.85\,meV/atom (vs.\ 1.69 for DeePMD), a force RMSE of 45.67\,meV/\AA\ (vs.\ 46.34), and a torque RMSE of 11.16\,meV/$\mu_{\rm B}$ (vs.\ 12.58).
Both models reach comparable accuracy across all three quantities, confirming that the lightweight NEP architecture does not sacrifice physical fidelity relative to the deep neural-network potential.

\begin{table}[htp]
\centering
\caption{Accuracy comparison between NEPSPIN and DeePMD on the FeGe spin--lattice validation set.
\textbf{Values are placeholders pending final validation.}}
\label{tab:accuracy_comparison}
\begin{tabular}{lccc}
\toprule
Property & Unit & NEPSPIN & DeePMD \\
\midrule
Energy RMSE     & meV/atom          & 1.85   & 1.69  \\
Force RMSE      & meV/\AA           & 45.67   & 46.34  \\
Magnetic Torque RMSE     & meV/$\mu_{\rm B}$ & 11.16   & 12.58  \\
\bottomrule
\end{tabular}
\end{table}

\begin{figure}[htp]
    \centering
    \includegraphics[width=0.95\linewidth]{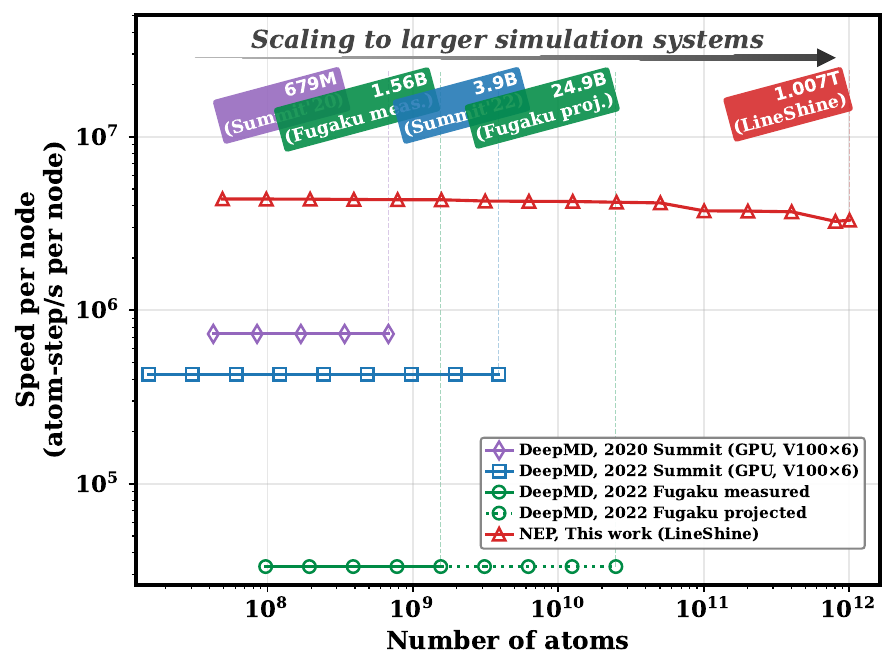}
%     \caption{ Weak-scaling throughput of the NEPSPIN inference engine on LineShine, compared with prior Gordon-Bell-class atom-only MLMD campaigns. This figure isolates throughput and accessible system size; coupled spin–lattice capability is compared separately in Table I. Weak-scaling performance of NEPSPIN spin-lattice dynamics on the LineShine
%     Weak-scaling performance of NEPSPIN spin-lattice dynamics on the LineShine
% supercomputer for a water system, benchmarked from 1 to 20{,}480 nodes
% (49{,}152{,}000 atoms per node).
% Per-node throughput (atom-step/s per node) as a function of system size,
% compared with DeepMD results on Summit
% (2020~\cite{jia2020pushing}; 2022~\cite{guo2022extending})
% and on Fugaku (2022~\cite{guo2022extending}, measured as solid line,
% projected to full machine as dotted line).
% Colored markers indicate the maximum atom count achieved by each campaign;
% the arrow illustrates the progressive expansion of simulation scale.}
\caption{Per-node throughput (atom-step/s) of NEPSPIN on LineShine (1--20{,}480 nodes, water system) compared with DeepMD on Summit~\cite{jia2020pushing} and Fugaku~\cite{guo2022extending}. Colored markers denote the maximum atom count of each campaign.}
    \label{fig:nep_vs_dp_speed}
    \vspace{-10pt}
\end{figure}

\label{sec:weak}
\begin{figure*}[!htp]
    \centering
    \includegraphics[width=0.98\textwidth]{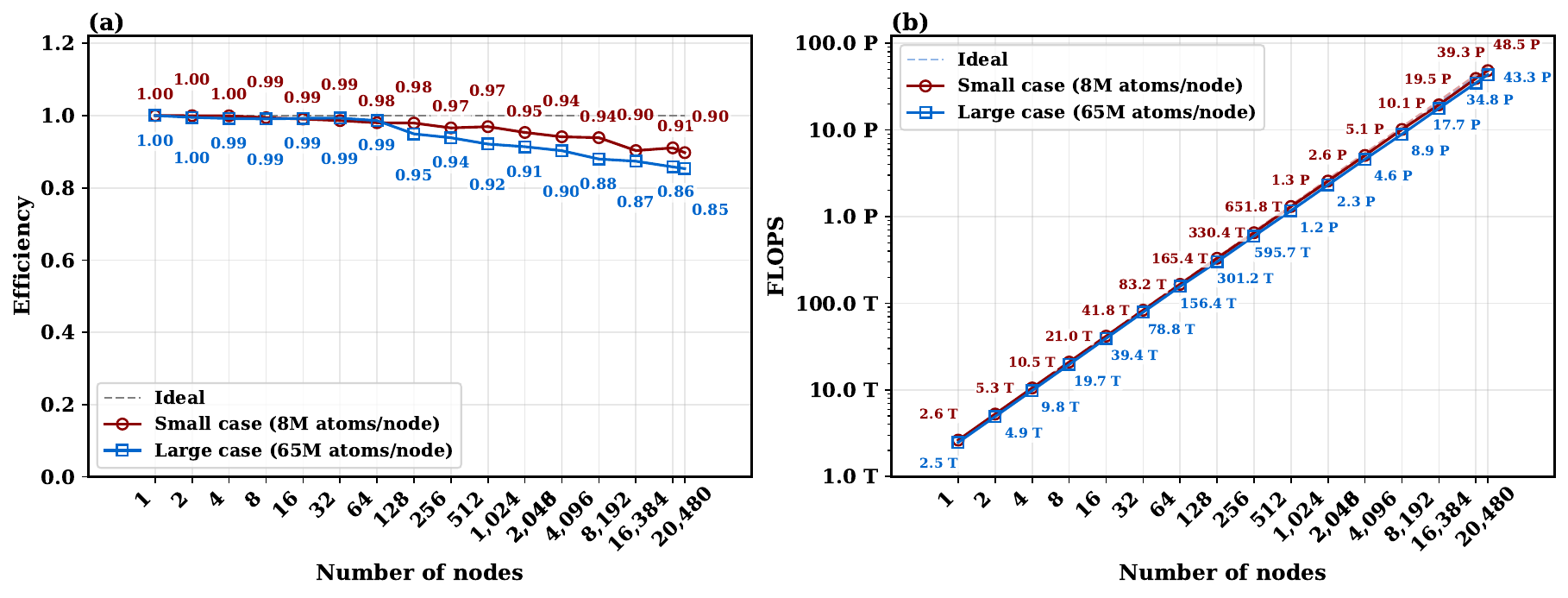}
    \caption{Weak-scaling results from 1 to 20{,}480 nodes.
    (a)~Parallel efficiency for the small case (8\,M~atoms/node) and large case (65\,M~atoms/node).
    (b)~Sustained application-level FLOPS for both configurations.
    Dashed lines indicate ideal scaling.}
    \label{fig:weak_scaling}
\end{figure*}

\begin{figure}[htp]
    \centering
\includegraphics[width=0.5\textwidth]{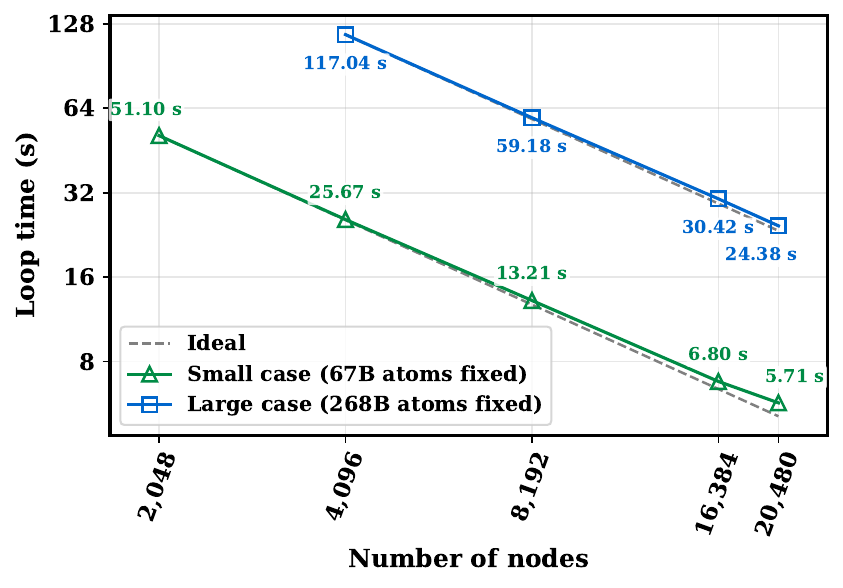}
    %\caption{Strong-scaling loop time for a fixed system of 67.1~billion atoms, from 2{,}048 to 20{,}480 nodes.The dashed line indicates ideal linear scaling from the 2{,}048-node baseline.}
    \caption{Strong-scaling loop time for two fixed-size systems.
    \textit{Green triangles}: $6.71\times10^{10}$~atoms (67.1~billion), scaled from
    2{,}048 to 20{,}480 nodes ($10\times$ range).
    \textit{Blue squares}: $2.68\times10^{11}$~atoms (268.4~billion), scaled from
    4{,}096 to 20{,}480 nodes ($5\times$ range).
    Dashed lines indicate ideal linear speedup from each respective baseline node count.}
    \label{fig:strong_scaling}
\end{figure}

%Fig. 4: 
%\paragraph{Throughput scaling with system size.}

%\noindent\textbf{
\textit{Throughput scaling with system size:}
%}
%We evaluate the weak-scaling performance of our NEPSPIN implementation on the LineShine supercomputer using a liquid water system with $49{,}152{,}000$ atoms per node. The total throughput scales from $4.38 \times 10^{6}$~atom-step/s on a single node to $6.77 \times 10^{10}$~atom-step/s on $20{,}480$ nodes, representing a $1.55 \times 10^{4}$-fold increase and corresponding to approximately $1.0 \times 10^{12}$ atoms simulated simultaneously. Compared to the DeepMD results on Fugaku~\cite{guo2022extending}, which achieved a maximum measured throughput of $3.29 \times 10^{8}$~atom-step/s at $1/16$ of the machine (${\sim}9{,}874$ nodes, $1.56 \times 10^{9}$ atoms), our implementation on LineShine surpasses this by more than two orders of magnitude in absolute throughput while extending the accessible system size from $1.6 \times 10^{9}$ to over $1.0 \times 10^{12}$ atoms.
To ensure a controlled comparison with the DeePMD-based Gordon Bell campaigns~\cite{jia2020pushing,guo2022extending}, which all adopted liquid water as their benchmark system, we reproduce the same water simulation on LineShine using our NEPSPIN engine with $49{,}152{,}000$ atoms per node.
This choice isolates differences in framework efficiency and hardware platform from those in model complexity or physical system properties.
The total throughput scales from $4.38 \times 10^{6}$~atom-step/s on a single node to $6.77 \times 10^{10}$~atom-step/s on $20{,}480$ nodes, representing a $1.55 \times 10^{4}$-fold increase and corresponding to approximately $1.0 \times 10^{12}$ atoms simulated simultaneously.
Compared to the DeepMD results on Fugaku~\cite{guo2022extending}, which achieved a maximum measured throughput of $3.29 \times 10^{8}$~atom-step/s at $1/16$ of the machine (${\sim}9{,}874$ nodes, $1.56 \times 10^{9}$ atoms), our implementation on LineShine surpasses this by more than two orders of magnitude in absolute throughput while extending the accessible system size from $1.6 \times 10^{9}$ to over $1.0 \times 10^{12}$ atoms.

As illustrated by the milestone markers in Figure~\ref{fig:nep_vs_dp_speed}, each successive Gordon Bell campaign has pushed the atom-count frontier---from $679$\,M (Summit~2020) to $3.9$\,B (Summit~2022) and $1.56$\,B measured on Fugaku~2022---while LineShine now reaches $1.0$\,T atoms.
On a per-node basis, LineShine delivers ${\sim}130\times$ the throughput of Fugaku and ${\sim}10\times$ that of Summit~2022, despite providing only ${\sim}2.8\times$ the FP64 peak per node.
This disproportionate advantage reflects the lower arithmetic intensity of the NEP descriptor relative to DeepMD, combined with the LX2 node's efficient utilization of SME/SVE vector units and on-package HBM for memory-bound neighbor-list operations.

% --- 7.2 Weak scaling: constant work per node, measures overhead growth
\subsection{Weak scaling}

In the weak-scaling study, the per-node workload is held constant while the number of nodes increases from 1 to 20{,}480.
Two representative configurations are evaluated: a \emph{small case} with $8.19\times10^{6}$~atoms per node and a \emph{large case} with $6.55\times10^{7}$~atoms per node ($8\times$ larger).
At full scale (20{,}480 nodes, 12.45\,M cores), the small case evolves $1.68\times10^{11}$~atoms (167.77~billion) and the large case evolves $1.34\times10^{12}$~atoms (1.34~trillion).

Figure~\ref{fig:weak_scaling}(a) reports the parallel efficiency.
At full machine scale (20{,}480~nodes), the small case retains 89.7\% efficiency and the large case 85.3\%.
The small case is consistently more efficient, owing to better cache reuse within the per-core L1/L2 hierarchy on the LX2 architecture.
The primary source of efficiency loss at extreme scale is the increasing surface-to-volume ratio of each MPI sub-domain, which raises the ghost-atom fraction and inter-node communication volume.
The high arithmetic intensity of the NEPSPIN angular kernels---$O(N_{\mathrm{neigh}}\times L_{\max}^2)$ FLOPs per atom versus $O(N_{\mathrm{local}}^{2/3})$ halo exchange---amortizes this overhead and maintains high efficiency across the full scaling range.
Figure~\ref{fig:weak_scaling}(b) shows the sustained application-level FLOPS.
Both configurations scale nearly linearly with node count, reaching 48.5\,PFLOPS (small case) and 43.3\,PFLOPS (large case) at 20{,}480 nodes.
The single-node baseline is 2.64\,TFLOPS (small) and 2.48\,TFLOPS (large), both in double precision.

% --- 7.3 Strong scaling: fixed total work, measures speedup and time-to-solution
\subsection{Strong scaling}
\label{sec:strong}

Strong-scaling behaviour is characterised with two fixed-size systems---67\,B and 268\,B atoms---spanning 2{,}048 to 20{,}480 nodes (Figure~\ref{fig:strong_scaling} and Table~\ref{tab:strong_scaling}).
Both cases exhibit near-ideal speedup at moderate concurrency: the first doubling of node count delivers $1.99\times$ and $1.98\times$ speedup, respectively.

\begin{figure*}[!htp]
    \centering
    \includegraphics[width=0.98\linewidth]{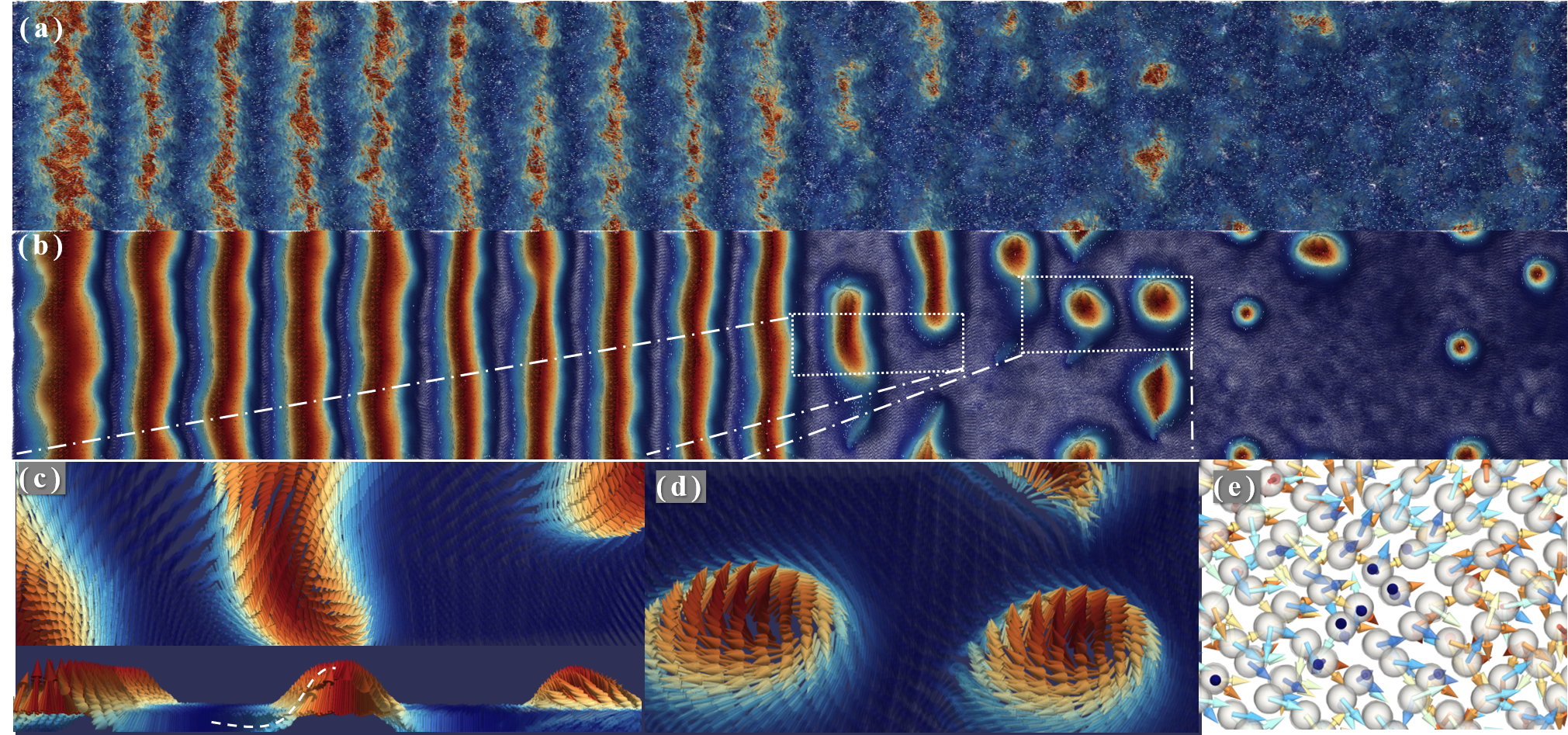}
    \caption{Representative subregion extracted from the full-machine extreme-scale spin--lattice simulation, illustrating the spatial evolution from helical order to skyrmionic textures in a magnetic stripe. The displayed region has a size of $1.141~\mathrm{\mu m}\times169.7~\mathrm{nm}\times1.8~\mathrm{nm}$ and is taken from the full production run under the same simulation protocol.  The system is initialized from a random spin distribution and evolved at $T=160~\mathrm{K}$ under a magnetic-field gradient along the $x$ direction, with field values of $0$, $0.05$, $0.1$, and $0.15~\mathrm{T}$. The upper panel (a) and (b) shows the full texture distribution at $T=160~\mathrm{K}$ and after-relaxation, (b) The lower-left panel enlarges a helical region, and the side view beneath it illustrates the spin rotation within the helix, with the white dashed lines tracing the helical winding.  The lower-center panel (d) enlarges a skyrmion region, where three skyrmions exhibit a lattice-like arrangement. The lower-right panel (e) enlarges a area where a locally disordered spin and atom configuration develops as the helical stripe breaks. 
    The figure highlights the ability of the present framework to capture both global texture reorganization and local topological spin structure at real temperature.
    }
    \label{fig:fig_total}
     \vspace{-14pt}
\end{figure*}

At full machine scale the 268\,B-atom case retains 96.0\% parallel efficiency ($4.80\times$ over 4{,}096 nodes), while the 67\,B-atom case reaches 89.6\% ($8.96\times$ over 2{,}048 nodes) despite a $10\times$ node-count range that reduces the per-node workload to only ${\sim}3.3$\,M atoms.
The efficiency gap between the two cases is consistent with the scaling characteristics of short-range force models: because the NEPSPIN force evaluation scales as $O(N_{\mathrm{local}}\cdot N_{\mathrm{neigh}})$ while ghost-atom exchange grows as $O(N_{\mathrm{local}}^{2/3})$, smaller sub-domains expose a larger communication fraction.
Nevertheless, the architecture-aware optimizations described in Section~\ref{sec:innovations} keep efficiency above 89\% across the entire scaling range.

\begin{table}[htp]
\centering
\caption{Strong-scaling results for two fixed-size systems.
         Speedup and efficiency are relative to the smallest node count
         used for each case (2{,}048 for the small case;
         4{,}096 for the large case).}
\label{tab:strong_scaling}
\begin{tabular}{clrcrc}
\toprule
Case & Nodes & Atoms/node & Loop time (s) & Speedup & Eff.\ (\%) \\
\midrule
\multirow{5}{*}{67\,B}
 &  2{,}048  & 32.77\,M & 51.10 & 1.00$\times$ & 100.0 \\
 &  4{,}096  & 16.38\,M & 25.67 & 1.99$\times$ &  99.5 \\
 &  8{,}192  &  8.19\,M & 13.21 & 3.87$\times$ &  96.7 \\
 & 16{,}384  &  4.10\,M &  6.80 & 7.52$\times$ &  93.9 \\
 & 20{,}480  &  3.28\,M &  5.71 & 8.96$\times$ &  89.6 \\
\midrule
\multirow{4}{*}{268\,B}
 &  4{,}096  & 65.54\,M & 117.04 & 1.00$\times$ & 100.0 \\
 &  8{,}192  & 32.77\,M &  59.18 & 1.98$\times$ &  98.9 \\
 & 16{,}384  & 16.38\,M &  30.42 & 3.85$\times$ &  96.2 \\
 & 20{,}480  & 13.11\,M &  24.38 & 4.80$\times$ &  96.0 \\
\bottomrule
\end{tabular}
\end{table}

\subsection{Sustained floating-point performance}
\label{sec:sustained}

At the full scale of 20{,}480 nodes (12.45\,M~cores), the application sustains 48.5\,PFLOPS (small case) and 43.3\,PFLOPS (large case) in double precision, as reported in Figure~\ref{fig:weak_scaling}(b).
The near-linear growth of sustained FLOPS with node count confirms that the computation-to-communication ratio remains favorable across the entire scaling range.
At single-node level, the small case achieves 2.64\, TFLOPS and the large case 2.48\, TFLOPS per node; at 20{,}480 nodes, these per node throughputs decrease to 2.37\, TFLOPS and 2.11\, TFLOPS respectively, consistent with the 89.7\% and 85.3\% parallel efficiencies reported above.
 Performance is limited primarily by the gradual increase in communication overhead rather than by load imbalance or memory bandwidth saturation, indicating that architecture-aware kernel design and SoA data layout effectively exploit the SVE2 and SME capabilities of the underlying hardware.

\textbf{All results are obtained in double precision because accurate spin--lattice dynamics---particularly the spin-orbit-mediated torque terms---requires full FP64 fidelity throughout the force and torque evaluation pipeline.}
At the deployed scale of 20{,}480 nodes (theoretical peak $120.6\,\text{TFLOP/s} \times 20{,}480 \approx 2.47\,\text{EFLOP/s}$), \textbf{the sustained 48.5\,PFLOPS corresponds to approximately 2.0\% of the FP64 peak}.
Low peak utilization is, in fact, characteristic of particle-based and neighbor-list-driven applications on modern hardware.
Fedeli et al.~\cite{fedeli2022pushing} report that electromagnetic particle-in-cell (PIC) codes---whose dominant kernels (current deposition and field gathering) share the same gather/scatter memory-access pattern as MLIP neighbor-list operations---sustain only 1--3\% of the DP peak on Fugaku A64FX CPUs, with the HPCG benchmark proposed as a more representative roofline reference than HPL.
They note that even the highly optimized VPIC code achieved only 13.2\% utilization on Roadrunner (2008), and that 2--10\% is the typical range for production PIC codes across architectures.
Our ${\sim}2\%$ utilization on the ARM-based LineShine is consistent with this picture: the NEPSPIN kernels are dominated by neighbor-list traversal, descriptor accumulation via irregular gather/scatter, and shallow-network GEMV, all of which are memory-bandwidth-bound and cannot saturate the SME matrix engine designed for dense outer-product GEMM.

By comparison, DeepMD-kit reports ${\sim}22\%$ utilization on both Fugaku and Summit~\cite{guo2022extending}, owing to its deep embedding networks whose inference is dominated by dense matrix multiplications with high arithmetic intensity.
However, the Fugaku figure of 119\,PFLOPS is a \emph{projected} estimate extrapolated from runs on $1/16$ of the machine, not a full-scale measurement.
NEP, by contrast, adopts a compact Chebyshev polynomial descriptor with $O(10^{3})$ trainable parameters---deliberately trading arithmetic density for per-atom efficiency.
This design delivers ${\sim}10$--$130\times$ higher per-node throughput than DeepMD (Section~\ref{sec:deepmd}) at comparable accuracy, making time-to-solution the more relevant metric for this class of applications.

%As shown in Section~\ref{sec:deepmd}, this design delivers ${\sim}10\times$ higher per-node throughput than DeepMD on Summit~2022 and ${\sim}130\times$ that on Fugaku, at comparable accuracy (Table~\ref{tab:accuracy_comparison}), making time-to-solution, rather than FLOPS utilization, the appropriate performance metric for this class of applications.

%This design yields a ${\sim}10\times$ lower time-to-solution per atom compared with DeepMD at comparable accuracy (Table~\ref{tab:accuracy_comparison}), making TtS, rather than FLOPS utilization, the appropriate performance metric for this class of applications.

% ====================================================================
% 8. IMPLICATIONS — GB requirement: 1 page max;
%    implications for future systems and applications
% ====================================================================
\section{Implications}
% \begin{figure}[htp]
%     \centering
%     %\includegraphics[width=0.95\linewidth]{Fig_nep_vs_dp_speed.pdf}
%     \fbox{\parbox{0.9\linewidth}{\centering\vspace{2cm}
%     \textcolor{red}{\textbf{Placeholder:} The coexistance of helix, cone and skyrmion's lattice}
%     \vspace{2cm}}}
%     \caption{}
%     \label{fig:nep_physics_application}
% \end{figure}

% --- Spin-lattice dynamics simulation of helix and skyrmion:
%     the new physics enabled by the extreme-scale capability

A flagship application enabled by the present framework is the real-temperature transformation from helical order to skyrmion textures in FeGe. At the same time, the micrometer-scale simulation domain accommodates many helical and skyrmionic periods within a single atomistically resolved calculation, providing access to competing boundaries, fluctuation-mediated nucleation channels, and texture reorganization pathways.

Under the thermal and field protocol described in the text, the system develops a spatially heterogeneous topological texture landscape in the stripe (Fig.~\ref{fig:fig_total}), ranging from multi-helix textures to isolated skyrmions and skyrmion-lattice-like regions. The enlarged local snapshot further shows that skyrmion nucleation proceeds through a localized rupture of the helical texture, accompanied by the formation of a strongly twisted and partially disordered intermediate region, which subsequently evolves into a topological seed. Importantly, this helix-breaking event is observed only when finite temperature is applied to both the lattice and spin degrees of freedom together with the external magnetic field. In contrast, under the same field but without thermal activation, the helical texture remains intact within the simulation time window and no skyrmion nucleation is observed. This indicates that the magnetic field alone is insufficient to overcome the topological and energetic barrier associated with helix breaking, whereas thermal fluctuations in the coupled spin--lattice system are essential for activating the transition pathway. 

More broadly, this work demonstrates that extreme-scale atomistic spin–lattice simulation can serve as a predictive tool for linking atomic structure, magnetic order, microstructure, and functional response in magnetic materials. This capability is directly relevant to topological-spin-texture-based spintronic devices, to microstructure- and defect-controlled magnetic performance, and to the broader class of advanced functional materials governed by coupled order parameters.

\section*{Acknowledgment}
This work was supported in part by the National Key R\&D Program of China under Grant No.\ 2025YFB3003603.
%The use of AI-assisted writing tools was employed to improve the clarity and language of portions of this manuscript. All scientific content, results, and conclusions are the sole responsibility of the authors.

% ====================================================================
% REFERENCES — GB requirement: 1 page max
% ====================================================================
\bibliographystyle{IEEEtran}
\bibliography{reference}

\end{document}